\documentclass{IOS-Book-Article}
\usepackage{mathptmx}
\usepackage{soul}\setuldepth{article}

\def\hb{\hbox to 10.7 cm{}}

\usepackage{cite}
\usepackage{amsmath,amssymb,amsfonts}
\usepackage{algorithmic}
\usepackage{graphicx}
\usepackage{subcaption}
\usepackage{textcomp}
\usepackage{xcolor}
\def\BibTeX{{\rm B\kern-.05em{\sc i\kern-.025em b}\kern-.08em
    T\kern-.1667em\lower.7ex\hbox{E}\kern-.125emX}}

\newtheorem{theorem}{Theorem}[section]

\newtheorem{proposition}[theorem]{Proposition}

\newtheorem{definition}[theorem]{Definition}
\newtheorem{remark}[theorem]{Remark}

\def\bZ{\mathbb{Z}}
\def\bR{\mathbb{R}}

\def\cD{\mathcal{D}}
\def\cE{\mathcal{E}}

\def\cM{\mathcal{M}}
\def\cN{\mathcal{N}}

\def\cP{\mathcal{P}}

\def\cR{\mathcal{R}}

\def\cV{\mathcal{V}}

\def\fC{\mathfrak{C}}

\def\fe{\mathfrak{e}}

\begin{document}

\pagestyle{headings}
\def\thepage{}

\begin{frontmatter}              

\title{Community detection, pattern recognition, and hypergraph-based learning: approaches using metric geometry and persistent homology}

\markboth{}{September 25, 2020\hb}
\author[A]{\fnms{Dong Quan Ngoc} \snm{Nguyen}
\thanks{Corresponding Author: Department of Applied and Computational Mathematics and Statistics, \\
University of Notre Dame, Notre Dame, Indiana 46556 USA; E-mail:
dongquan.ngoc.nguyen@nd.edu}},
\author[A]{\fnms{Lin} \snm{Xing}},
and
\author[A]{\fnms{Lizhen} \snm{Lin}}

\runningauthor{D.Q.N. Nguyen et al.}
\address[A]{Department of Applied and Computational Mathematics and Statistics, \\
University of Notre Dame, \\
Notre Dame, Indiana 46556 USA}

\begin{abstract}
Hypergraph data appear and are hidden in many places in the modern age. They are data structure that can be used to model many real data examples since their structures contain information about higher order relations among data points. One of the main contributions of our paper is to introduce a new topological structure to hypergraph data which bears a resemblance to a usual metric space structure. Using this new topological space structure of hypergraph data, we propose several approaches to study community detection problem, detecting persistent features arising from homological structure of hypergraph data. Also based on the topological space structure of hypergraph data introduced in our paper, we introduce a modified nearest neighbors methods which is a generalization of the classical nearest neighbors methods from machine learning. Our modified nearest neighbors methods have an advantage of being very flexible and applicable even for discrete structures as in hypergraphs. We then apply our modified nearest neighbors methods to study sign prediction problem in hypegraph data constructed using our method.

\end{abstract}

\end{frontmatter}
\markboth{September 7, 2020\hb}{September 7, 2020\hb}

\section{Introduction}
One of the challenges in the modern age is to classify data arising from many resources; for example, following the rapid developments of several areas in mathematics, a large number of publications in mathematics creates a tremendous amount of data, which signifies useful information such as relationships (or collaborations) among authors and their publications, and their influences on development of mathematics. It is often the case that analyzing such data is not straightforward, and very difficult task because of the extremely fast growth of relations among data, and of data itself.

In this paper, we propose several methods to analyze \textit{hypergraph data}. Recall that a hypergraph $X$ is a pair $(\cV(X), \cE(X))$, where $\cV(X)$ is the set of data points (called vertices of $X$), and $\cE(X)$ is a subset of the power set of $\cV(X)$ which represents relations among data points. Each element in $\cE(X)$ is called a \textit{hyperedge}. Note that by abuse of notation, we sometime use the same symbol for $X$ and its set of vertices. 

A standard example of a hypergraph is a collaboration network in which the set of vertices consists of mathematicians, and a group of mathematicians (not necessarily only two) forms a hyperedge if they have at least one joint publication. Many real data can be modeled as a hypergraph. Applications of hypergraph data are diverse such as in protein function prediction (see \cite{THK}), and other areas (for example, see \cite{Hyper1}, \cite{Hyper2}, \cite{Hyper3}, \cite{Hyper4}). 

The aim of this paper is to propose several approaches to studying community detection, pattern recognition, and sign prediction problem. Our approaches use main tools from metric geometry (see, for example, \cite{BBI}), combined with techniques from geometric and topological inference, to adapt classical techniques such as nearest neighbors methods into the hypergraph settings. More precise, for a given hypergraph data, we introduce a class of metrics modulo certain equivalence relations (for a precise definition, see Section \ref{S-notions}) to equip such hypergraph with a metric space structure. Using these structures, we propose several approaches to detect features from hypergraphs; for example, only using distance matrix approach, we provide a way to approach to community detection problem. Based on the metric space structure, we apply tools from algebraic topology to propose a method for detecting \textit{persistent features} arising from homological structures hidden in a hypergraph. Such approach has an advantage of visualization of the space structure of data which provides a visual insight into the topological structure of hypegraph data. Also based on the metric space structure of hypergraph data, we introduce a modified nearest neighbors method which is a generalization of the classical nearest neighbors method from machine learning. Using our modified nearest neighbors methods, we apply to sign prediction problem on hypegraph data constructed by our method. 

One of the novel and main features in our paper is that we propose a new type of hypergraph data (which we coin the term ``\textit{congruence hypergraph data}'') which are very easy to construct and implement, and very flexible for testing our theories. 

The structure of our paper is as follows. In Section \ref{S-notions}, we introduce several notions and our main metric on hypergraphs that will be used throughout the paper. In Section \ref{S-Analysis-hypergraph}, we introduce congruence hypergraph data, and several methods for analyzing hypergraph data including the distance matrix approach, homology-based learning, and modified nearest neighbors methods. Several examples will be performed on congruence hypergraph data which we introduce in Subsection \ref{SubS-example}.

\section{Basic notions}
\label{S-notions}

\subsection{Metrics modulo equivalence relations}

Let $X$ be a set. An \textit{equivalence relation}, denoted by $\cong$, on $X$ is a subset of $X \times X$ such that the following conditions are true: 
\begin{itemize}

\item [(i)] (\textbf{Reflexivity}) $(a, a) \in \; \cong$ for every $a \in X$.

\item [(ii)] (\textbf{Symmetry}) $(a, b) \in \; \cong$ if and only if $(b, a) \in \;\cong$.

\item [(iii)] (\textbf{Transitivity}) if $(a, b) \in \; \cong$ and $(b, c) \in \;\cong$ then $(a, c) \in \; \cong$. 

\end{itemize} 

When $(a, b) \in \; \cong$, we say that $a$ is $\cong$--equivalent to $b$. Throughout this paper, in order to signify this relation, we write $a \cong b$ whenever $(a, b) \in \; \cong$. 

For a given high order network (which is another terminology for hypergraph data), one of the problems that we address in this paper is concerned with \textit{distinguishing communities} in the network. It is clear that there are many examples of networks in which several communities are viewed as identical communities with respect to certain properties that one wants to know about these networks. So if we view a given high order network $X$ as a hypergraph, in order to use a metric geometry approach to the community detection problem, it is natural to introduce a metric (or distance) on $X$ modulo a certain equivalence relation which will be explicitly introduced depending on the structure of $X$. Before making it clear what exactly we mean by this point of view, using an example of high order collaboration network, we first introduce the notion of a metric modulo an equivalence relation.
\begin{definition}
Let $X$ be a set, and let $\cong$ be an equivalence relation on $X$. A mapping $d : X \times X \to \bR$ is said to be a metric on $X$ modulo the equivalence relation $\cong$ if the following conditions are satisfied:
\begin{itemize}

\item[(i)] $d(a, b) \ge 0$ for all $a, b \in X$.

\item [(ii)] $d(a, b) = 0$ if and only if $a \cong b$.

\item [(iii)] (\textbf{Symmetry}) $d(a, b) = d(b, a)$ for all $a, b \in X$.

\item [(iv)] (\textbf{Triangle inequality}) for any $a, b ,c \in X$, 
$$d(a, b) \le d(a, c) + d(c, b).$$

\end{itemize}

\end{definition}

A set $X$ equipped with a metric modulo an equivalence relation $\cong$, say  $d : X \times X \to \bR$ is called a metric space modulo $\cong$ . In notation, we write $(X, d)$ to indicate this metric space modulo $\cong$.

\subsection{Hypergraphs equipped with intrinsic properties}
\label{SubS-Main-Metric}

Let $X$ be a set. In order to create a \textit{hypergraph} structure on $X$, we view the set of all points in $X$ as the set of vertices $\cV(X)$, and one needs to identify the relations among points in $X$, which one can view as the set of hyperedges of $X$, denoted as $\cE(X)$. A hyperedge having exactly $\ell$ vertices is called an $\ell$-hyperedge. The way which one identifies hyperedges in $X$, signifies certain properties pertained to the set $X$ that we want to study. For example, let $X$ be a set of mathematicians. In order to study how collaborative the mathematicians in $X$ are, we introduce a hypergraph structure on $X$ as follows. The set of vertices of $X$ simply consists of all mathematicians in $X$. A group of mathematicians, say $m_1,\ldots, m_{\ell}$ in $X$ forms an $\ell$-hyperedge if they have at least one joint publication. In this way, the set $X$ becomes a hypergraph in which the construction of hyperedges signifies the collaboration among mathematicians in $X$. 

In many real data examples, one is not only interested in the hypergraph structure of $X$, but also in knowing certain properties attached to such structure but hidden in the hyperedge data. For example, in the collaboration network just described, in order to study in which areas of mathematics the mathematicians in $X$ have joint publications, we can associate to each hyperedge the \textit{main area of mathematics} in which the joint publication of the hyperedge belongs. If an $\ell$-hyperedge $\fe$ is formed out of the joint paper, say $P$, of $\ell$ mathematicians $m_1, \ldots, m_{\ell}$, and the paper $P$ is mainly concerned about number theory, then one can define $\Gamma(\fe) = \text{number theory}$. Hence one obtains a mapping $\Gamma$ from the set of hyperedges of $X$ to the set of all areas in mathematics. Studying such a map $\Gamma$ provides insight into the relationships between joint publications of mathematicians in $X$ and their contributions to certain fields in mathematics. Motivated by this example in mind, we introduce a notion of hypergraphs equipped with certain properties.

\begin{definition}
Let $X = (\cV(X), \cE(X))$ be a hypergraph, and let $\cP$ be a nonempty set. The hypergraph $X$ is called a \textit{hypergraph equipped with properties $\cP$} if there is a map $\Gamma : \cE(X) \to \cP$ which associates each hyperedge in $X$ to a unique element in $\cP$. 

In notation, we write $\{X, \cP\}_{\Gamma}$ to indicate $X$ is a hypergraph equipped with properties $\cP$, where the subscript $\Gamma$ is a map from $\cE(X)$ to $\cP$.

\end{definition}

In this subsection, for each hypergraph equipped with properties $\cP$, say $\{X, \cP\}_{\Gamma}$, we introduce a metric space structure on $X$, which provides a way to distinguishing communities in $X$. We begin by defining a notion of neighborhood of a vertex which is more suitable for defining a metric on the hypergraph $X$. 
\begin{definition}
\label{neighborhood-def}
Let $X = (\cV(X), \cE(X))$ be a hypergraph. Let $a$ be a vertex in $X$. The neighborhood of $a$, denoted by $\cN(a)$, is defined by
\begin{align*}
\cN(a) = \{a\} \cup \{b \in \cV(X) \; | \; \text{$\exists \fe \in \cE(X)$ such that $\{a, b\} \subset \fe$ }\}.
\end{align*}

\end{definition}

The next result is clear from the above definition.

\begin{proposition}
Let $X$ be a hypergraph whose set of vertices consists of $a_1, \ldots, a_n$. Then $X$ can be decomposed into $n$ neighborhoods, say $\cN(a_1), \ldots, \cN(a_n)$ of the form
$$X = \cN(a_1) \cup \cdots \cup \cN(a_n).$$

\end{proposition}

\begin{remark}\label{Rem1}
For the community detection problem, the proposition above plays a key role. Indeed, by communities in $X$, we mean neighborhoods of each vertex. And thus in order to point out differences among communities, we are interested in finding out the exact differences among the populations of hyperegdes with specific properties in $\cP$, contained in these neighborhoods; more precisely, letting a property $P$ range over the set $\cP$, the differences between the neighborhoods of $a_i$ and $a_j$ are reflected in terms of the differences between the numbers of hyperedges contained in the neighborhoods of $a_i$ and $a_j$ whose values under the map $\Gamma$ is exactly $P$, i.e., they share the same property $P$. Because of this observation and the proposition above, we want to study neighborhoods of vertices in $X$ instead of the vertices themselves, and thus one views $X$ as a space whose \textit{points} are neighborhoods $\cN(a_i)$. So each neighborhood is in fact viewed as a \textit{single point} in the space $X$. 

The metric geometry approach we use in this paper is that we want to construct a metric $d$ on such a space $X$ which should incorporate information about the number of hyperedges in $X$ with specific properties $P$. And once such a metric is established for the space $X$, two neighborhoods $\cN(a_i), \cN(a_j)$ (i.e., two points in $X$) are different if and only if $d(\cN(a_i), \cN(a_j))$ is nonzero. And this is our first method for distinguishing communities in hypergraphs.

\end{remark}

Let $\{X, \cP\}_{\Gamma}$ be a hypergraph equipped with properties $\cP$. Suppose that the set of vertices in $X$ consists of $a_1, \ldots, a_n$, and the largest size of hyperedges in $X$ is $\ell$. Let $1 \le i \le n$ be an integer. For each property $P \in \cP$, let $\fC^1_P(a_i)$ be the number of $1$-hyperedges in $\cN(a_i)$ whose value under the map $\Gamma$ is $P$. In a similar manner, let $\fC^2_P(a_i)$ be the number of $2$-hyperedges in $\cN(a_i)$ whose value under the map $\Gamma$ is $P$.  In general, for any integer $1 \le m \le \ell$, let $\fC^m_P(a_i)$ be the number of $m$-hyperedges in $\cN(a_i)$ whose value under the map $\Gamma$ is $P$. Thus one obtains a unique double sequence $\left((\fC^m_P(a_i))_{1 \le m \le \ell}\right)_{P \in \cP}$ of non-negative real numbers for each neighborhood $\cN(a_i)$. 

We introduce an equivalence relation on the space $X$ which allows to identify certain points in $X$. Note that if two points, say $\cN(a_i)$ and $\cN(a_j)$ have the same double sequence $\left((\fC^m_P(a_i))_{1 \le m \le \ell}\right)_{P \in \cP} = \left((\fC^m_P(a_j)_{1 \le m \le \ell}\right)_{P \in \cP}$, then it is natural to view both of them as \textit{identical points} in $X$ since their hyperdege structures are exactly the same with respect to the map $\Gamma$ and the properties $\cP$. Hence it is natural to define a binary relation on $X$ as follows:  \textit{two points $\cN(a_i)$ and $\cN(a_j)$  are \textit{equivalent}, denoted by $\cN(a_i) \cong \cN(a_j)$ if their associated sequences $\left((\fC^m_P(a_i))_{1 \le m \le \ell}\right)_{P \in \cP}$, $\left((\fC^m_P(a_j))_{1 \le m \le \ell}\right)_{P \in \cP}$ are identical, i.e., $\fC^m_P(a_i)  = \fC^m_P(a_j)$ for all $1 \le m \le \ell$ and all $P \in \cP$}. One obtains the following.

\begin{proposition}
The binary relation ``$\cong$'' is an equivalence relation. 

\end{proposition}

For the rest of this paper, whenever we use the symbol $\cong$ on hypergraphs, we mean the equivalence relation ``$\cong$'' in the proposition above. 

Now we define a mapping $\cD : X \times X \to \bR_{\ge 0}$ as follows. For $1 \le i, j \le n$, define
\begin{align*}
\cD(\cN(a_i), \cN(a_j)) := \sum_{P \in \cP}\sum_{m = 1}^{\ell}|\fC^m_P(a_i) - \fC^m_P(a_j)|.
\end{align*}  

From the above definition, we obtain the following.

\begin{theorem}
\label{main-thm1}
The mapping $\cD$ defined above is a metric on $X$ modulo the equivalence relation $\cong$.

\end{theorem}

The proof of the above theorem will be given in the appendix. 

\begin{remark}

In \cite[Definition 7, p.1060]{LTVT}, Leontjeva et al. constructed a distance function (or metric) between hypergraphs which uses the sizes of hyperedges in hypergraphs, in contrast to our construction using the number of hyperedges of each size. Note that in \cite{LTVT}, Leontjeva et al. claimed their metric is the metric in the usual sense which is not correct. It is in fact a metric modulo an equivalence relation. 

\end{remark}

\section{Analysis for hypergraph data}
\label{S-Analysis-hypergraph}

In this section, we propose several methods for analyzing hypergraph data.  Instead of using real data examples as in most papers studying data structures in literature, we introduce in this paper a new type of data which is inspired from elementary number theory (or more precisely from the theory of congruences in number theory), and is extremely easy to construct. There are many advantages of using such data which can also be viewed as hypergraphs. For simplicity, we call such data \textit{congruence hypergraph data}. Firstly, these data are very easy to construct, simply using congruences in the ring of integers $\bZ$. Secondly, congruence hypergraph data are very \textit{diverse} and \textit{random}, which provide a reasonably fine data to immediately test theory without referring to other data resources which in turn take a huge amount of time to build. The \textit{randomness} of congruence hypergraph data allows to justify with high probability that any theory used to successfully test on such data can be also applied to real data examples. Lastly, for congruence hypergraph data, we can easily control the size of data. On letting the data size go to infinity, one can detect patterns hidden in the data which are often not available and straightforward if the data size is only limited to be finite. 

\subsection{Congruence hypergraph data}
\label{SubS-congruence-hypergraph}

We now describe congruence hypergraph data which relies on the theory of congruences in the ring of integers $\bZ$. 

Let $n$ be a positive integer, and $\{m_1, \ldots, m_n\}$ be a collection of positive integers. Take $n$ collections of integers, say $\{a_{i, 1}, \ldots, a_{i, m_i}\}$ for each $1 \le i \le n$ such that
\begin{align*}
\{a_{i, 1}, \ldots, a_{i, m_i}\} \bigcap \{a_{j, 1}, \ldots, a_{j, m_j}\} = \emptyset
\end{align*}
for any $i \ne j$.

 Consider $n$ sets of integers, say $V_i = \{a_{i, 1}, \ldots, a_{i, m_i}\}$ for each $1 \le i \le n$ so that $\#V_i = m_i$. We want to introduce a hypergraph structure on each $V_i$, and thus the set $X = V_1 \cup V_2 \cup \cdots \cup V_n$ becomes a hypergraph which is a disjoint union of subhypergraphs $V_i$.

Now take an integer $1 \le i \le n$. Let $s_i$ be an integer such that $2 \le s_i \le m_i$. We want to introduce a hypergraph structure on $V_i$ such that the largest size of hyperedges in $V_i$ is $s_i$.

Let $\{p_{2, i}, \ldots, p_{s_i, i}\}$ be a sequence of integers such that the $p_{j, i}$ are $\ge 2$ and not necessarily distinct. Correspondingly we choose a sequence of finite sets of integers $\{S_{2, i}, \ldots, S_{s_i, i}\}$ for each $1 \le i \le r$. 

Let $k$ be an integer such that $2 \le k \le s_i$. A $k$-tuple of integers $\{\alpha_1, \ldots, \alpha_k\}$ in $V_i$ forms a $k$-hyperedge if the following conditions are satisfied:
\begin{itemize}

\item [(i)] $\alpha_s - \alpha_r \equiv 0 \pmod{p_{k. i}}$ for any $1 \le s, r \le k$.

\item [(ii)] $\alpha_s \pmod{p_{k, i}}$ belongs in $S_{k, i}$ for any $1 \le s \le k$.

\end{itemize}

So we have obtained a subhypergraph structure for each of the $V_i$, and thus $X = \cup_{i = 1}^n V_i$ is a hypergraph which splits into disjoint subhypergraphs. Note that $X$ has exactly $m_1 + m_2 + \cdots + m_n$ vertices. 

\subsection{Main example}
\label{SubS-example}

The hypergrah data we use to test our proposed methods in this paper is motivated from the construction of congruence hypergraph data in Subsection \ref{SubS-congruence-hypergraph}. We now describe two hypergraphs that we use throughout this work. 

\subsubsection{First example}
\label{example1}
Let $X = \{1, \ldots, 1000\}$. We introduce a hypergraph structure on $X$ as follows. A pair $\{a, b\}$ in $X$ forms a $2$-hyperedge if and only if either $a, b \equiv 1 \pmod{2}$ or $a, b \equiv 0 \pmod{2}$. In other words, $a, b$ have the same parity. Now for each $3 \le n \le 9$, an $n$-tuple $\{a_1, \ldots, a_n\}$ forms an $n$-hyperedge if and only if 
\begin{align*}
a_i \pmod{n} =
\begin{cases}
0 \; \; \; &\text{if $n \equiv 0 \pmod{3}$} \\
1 \; \; \; &\text{if $n \equiv 1 \pmod{3}$} \\
2  \; \; \; &\text{if $n \equiv 2 \pmod{3}$}
\end{cases}
\end{align*}
for every $1 \le i \le n$.

Since this data is about integers, we are interested in properties regarding integers such as divisibility. For this reason, we  study, for example, the divisibility by $11$ of each vertex in a hyperedge in $X$. So it is natural to define a map $\Gamma : \cE(X) \to \{0, 1\}$ by letting, for each $n$-hyperedge $\{a_1, \ldots, a_n\}$ in $\cE(X)$,
\begin{align*}
\Gamma\left(\{a_1, \ldots, a_n\}\right) = 1
\end{align*}
if 
\begin{align}
\label{example-1st-metric-condition}
a_i \equiv 0 \pmod{11}
\end{align}
for every $1 \le i \le n$, and 
\begin{align*}
\Gamma\left(\{a_1, \ldots, a_n\}\right) = 0
\end{align*}
if condition (\ref{example-1st-metric-condition}) is not satisfied. 

The hypergraph $X$ above has very large number of hyperedges. Up to our knowledge, comparing with real data examples in literature, the hypergraph data $X$ above contains the \textit{largest} number of hyperedges which is very suitable for testing theories. For example, the number of hyperedges in the neighborhood (or community) of the vertex $1$ is approximately $2.3685 \times 10^{11}$. 

\subsubsection{Second example}
\label{example2}

Let $X$ be a set of integers obtained by randomly selecting $5000$ positive integers. The sizes of hyperedges range from $2$ to $9$. We randomly select $8$ integers, say $\{\alpha_2, \ldots, \alpha_9\}$, such that any two of them have no common divisors. The set of vertices $X$ are sorted in increasing order. For each $2 \leq n \leq 9$, we divide $X$ into $n$ subsets. The first subset, say $X_1$, contains vertices $a$ in $X$ such that $\min(X) \leq a \leq p_1$, where $\min(X)$ is the minimum value of $X$ and $p_1$ is the $1/n$-th percentile of $X$. For each $2 \le j \le n$, the $j$-th subset, say $X_j$, contains vertices $a$ in $X$ such that $p_j < a \le p_{j+1} $, where $p_j$ is the $j/n$-th percentile of $X$. An $n$-tuple $\{a_1, \ldots, a_n\}$ forms an $n$-hyperedge if and only if 
\begin{align*}
a_j \equiv 1 \pmod{\alpha_n} \; \text{and} \; a_j \in X_j
\end{align*} 
for every $1 \le j \le n$.

Then we randomly select an odd prime number $\beta$ that does not divide any elements in $\{\alpha_2, \ldots, \alpha_9\}$. The congruence classes modulo $\beta$ are divided into two sets, the first of which consists of $\{-(\beta - 1)/2, -(\beta - 1)/2 + 1, \ldots, -1, 0\}$, and the second of which consists of $\{1, 2, \ldots, (\beta - 1)/2\}$. Set
\begin{align*}
S^{-}_{\beta} = \{-(\beta - 1)/2, -(\beta - 1)/2 + 1, \ldots, -1, 0\},
\end{align*} 
and 
\begin{align*}
S^{+}_{\beta} =\{1, 2, \ldots, (\beta - 1)/2\}.
\end{align*}  

The properties of hyperedges are defined as follows. For each $n$-hyperedge $\{a_1, \ldots, a_n\}$ in $\cE(X)$,
\begin{align*}
\Gamma\left(\{a_1, \ldots, a_n\}\right) = -1
\end{align*}
\label{example-2nd-metric-condition}
if every $a_i$ modulo $\beta$ belongs to $S^{-}_{\beta}$, and
\begin{align*}
\Gamma\left(\{a_1, \ldots, a_n\}\right) = 1
\end{align*}
otherwise. 

\subsection{Using patterns from the distance matrices to recognize patterns in hypergraph data}
\label{SubS-patterns}

In this subsection, we describe a simple but useful approach to detecting communities in hypergraphs. Using this approach, one can identify which communities in a hypergraph are the same with respect to the equivalence relation ``$\cong$'' and the metric $\cD$ in Subsection \ref{SubS-Main-Metric}. On the other hand, one can also find patterns among vertices whose neighborhoods (i.e., communities) are identified as the same. 

Let $\{X, \cP\}_{\Gamma}$ be a hypergraph equipped with properties $\cP$. Assume that the set of vertices in $X$ consists of $a_1, \ldots, a_n$. Hence there are exactly $n$ neighborhoods (or communities), say $\cN(a_1), \ldots, \cN(a_n)$ which as remarked in Remark \ref{Rem1} can be viewed as points in the space $X$. Using the metric $\cD$, we equipped $X$ with a metric space structure in which each community $\cN(a_i)$ is a point of the metric space $X$. Since there are exactly $n$ points $\cN(a_1), \ldots, \cN(a_n)$ in the metric space, one obtains the \textit{distance matrix} of the finite metric space $X$, say $\cM_X$ of dimensions $n \times n$ of the form
$$\cM_X = \left(\cD(\cN(a_i), \cN(a_j))\right)_{1 \le i, j \le n},$$
where the $(i, j)$-entry in this matrix is the value $\cD(\cN(a_i), \cN(a_j))$. 

In order to identify which communities are the same in the hypergraph $X$, we identify all zero entries in $\cM_X$ except the diagonal. More precisely, let $1 \le i \le n$, and consider the $i$-th column in $\cM_X$. Define
$$Z_i = \{1 \le j \le n \; | \; \text{$j \ne i$ and $\cD(\cN(a_i), \cN(a_j)) = 0$}\}.$$
Then the set $Z_i$ consists of all vertices $j$ whose communities $\cN(a_j)$ are considered to be the same as the community $\cN(a_i)$. It is often the case that one can find patterns to describe $Z_i$. 

We use the hypergraph data in section \ref{example1}. In this case $X$ is a hypergraph whose vertices are $1, 2, \ldots, 1000$. Thus the distance matrix $\cM_X$ is of dimensions $1000 \times 1000$. For example, considering  the 1st column of $\cM_X$, we see that $Z_1$ contains exactly the following vertices: $29$, $43$, $71$,  $85$,  $113$,  $155$, $169$, $211$, $239$, $253$, $281$, $295$, $323$, $365$, $379$, $421$, $449$, $463$, $491$, $505$, $533$, $575$, $589$, $631$, $659$, $673$, $701$, $715$, $743$, $785$, $799$, $841$, $869$, $883$, $911$, $925$, $953$, $995$. And thus the communities (or neighborhoods) of these vertices are viewed as the same as that of the vertex $1$.

From the list of vertices in $Z_1$, one can recognize the patterns shared by the vertices in $Z_1$. Indeed all vertices $j$ in $Z_1$ satisfy the following four conditions: (i) $j \not\equiv 0 \pmod{3}$; (ii) $j \not\equiv 2 \pmod{5}$; (iii) $j \equiv \pm 1 \pmod{4}$; and (iv) $j \equiv 1 \pmod{7}$. 

From the distance matrix $\cM_X$, one also can identify the set of all distinct communities in $X$ consisting of the neighborhoods of $1$, $2$, $3$, $4$, $5$, $6$, $7$, $8$, $12$, $15$, $22$, 27$, 36$, $57$, $127$, and $162$ such that every community in $X$ is equal to exactly one of these neighborhoods.

\subsection{Homology-based learning using the metric $\cD$}
\label{homology using metric}
In this subsection, we use the \textit{persistent homology} of filtrations of simplicial complexes arising from a finite metric space modulo an equivalence relation $\cong$ $X$ to study the community detection problem. For simplicity, in this subsection, we simply call $X$ a metric space instead of a metric space modulo $\cong$. Let $\cV = \{a_1, \ldots, a_n\}$ be a finite set. A \textit{simplical complex} $X$ with vertex set $\cV$ is a set of finite subsets of $\cV$ satisfying the following conditions:
\begin{itemize}

\item [(i)] every element in $\cV$ belongs to $X$;

\item [(ii)] if $\tau \in X$ and $\sigma \subseteq \tau$, then $\sigma \in X$.

\end{itemize}

The elements of $X$ are called the \textit{simplices} of $X$. If a simplex $\sigma$ has exactly $k + 1$ elements, the \textit{dimension} of $\sigma$ is $k$, and we call $\sigma$ a $k$-simplex. 

To each simplicial complex $X$ one can associate a unique sequence of \textit{homology groups} $(H_k(X))_{k \ge 0}$ which contains information about topological and geometric properties of $X$. (See, for example, \cite{EH} or \cite{Zhu} for a notion of homology groups and their properties.) 

Now we describe how to use homology groups to identify distinct communities in hypergraphs. Let $\{X, \cP\}_{\Gamma}$ be a hypergraph equipped with properties $\cP$, and suppose that the set of vertices of $X$ consists of the vertices $a_1, \ldots, a_n$. We equip $X$ with the metric $\cD$ in Subsection \ref{SubS-Main-Metric}.

We introduce a method to attach to the finite metric space $X$ a collection of simplexes which one in turn can obtain the corresponding \textit{persistent homology sequences} and their \textit{barcodes} (see \cite{BCY}, \cite{code1}, \cite{code2}, \cite{code3}, \cite{code4} for persistent homology and barcodes.) We first recall a notion of Vietoris--Rips complex. 
\begin{definition}
Let $\epsilon > 0$, and let $X$ be a finite metric space with metric $\cD$. The Vietoris--Rip complex, denoted by $\cV\cR(X, \epsilon)$ is defined by the following condition: \textit{a $k + 1$-tuple $\{x_0, \ldots, x_{k}\}$ forms a $k$-simplex in $\cV\cR(X, \epsilon)$ if and only if $\cD(x_i, x_j) \le \epsilon$ for all $i, j$.}

\end{definition}

Let $h$ be a sufficiently large positive integer, and let $1 \le n_1 < n_2 < \cdots < n_h = n$ be a collection of positive integers. For each $1 \le k \le h$, define 
$$X_k = \{a_1, \ldots, a_{n_k}\}.$$
Note that $X_k \subset X$ for all $1 \le k \le h$, and thus each $X_k$ is a metric space with the same metric $\cD$. We have a filtration of metric spaces
$$X_1 \subset X_2 \subset \cdots \subset X_h = X.$$

For each finite metric space $X_k$, one obtains a filtration of Vietoris--Rip complexes $\cV\cR(X_k)$ from which one obtains the \textit{barcode} containing the topological and geometric information about $X_k$. The key observation using homology-based learning is that when $k$ ranges from $1$ to $h$, the barcodes of dimension $0$ will stabilize to have exactly $m$ bars, which signifies that \textit{there are exactly $m$ distinct communities} in the hypergraph $X$. Furthermore when considering the barcodes of dimension $1$, they will stabilize to have very similar forms when $k$ approaches to $h$. 

We illustrate the above method by testing this theory on  sub-hypergraph datasets of the congruence hypergraphs defined in the first and second examples in Subsections \ref{example1} and \ref{example2}. Note that for computing barcodes, we use the package \textbf{TDAstats} in \textit{R} (see \cite{TDA}). For the first example, let $h = 3$, and for each $0 \le k \le h$, set
$$X_k = \{1, \ldots, (2k + 1)100\},$$
and $X_3 = \{1, \ldots, 700\}$. One obtains exactly $4$ barcodes, each of which corresponds to exactly one $X_k$.

\begin{figure}
\centering
\begin{subfigure}[b]{.45\linewidth}
\includegraphics[width=\linewidth]{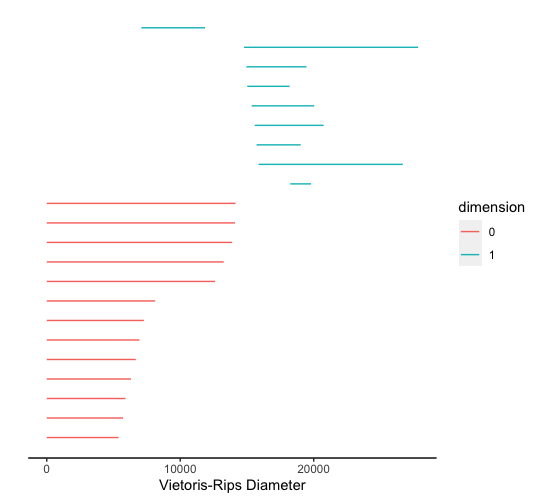}
\caption{Barcode of $X_0$.}\label{figX0}
\end{subfigure}
\begin{subfigure}[b]{.45\linewidth}
\includegraphics[width=\linewidth]{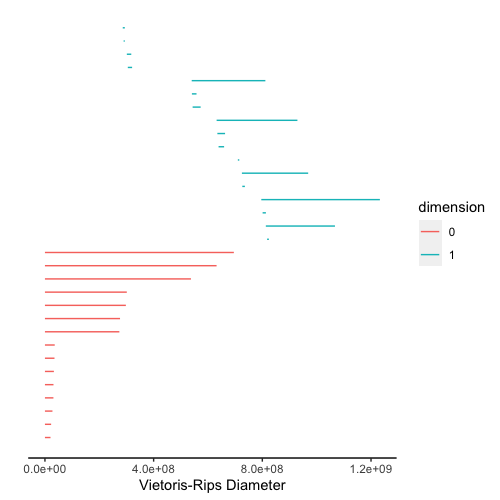}
\caption{Barcode of $X_1$.}\label{figX1}
\end{subfigure}

\begin{subfigure}[b]{.45\linewidth}
\includegraphics[width=\linewidth]{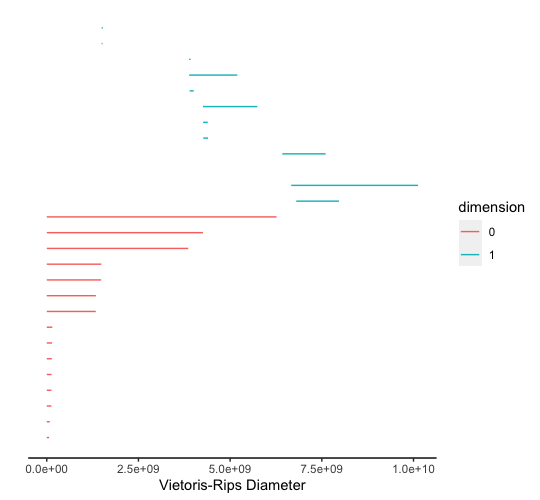}
\caption{Barcode of $X_2$.}\label{figX2}
\end{subfigure}
\begin{subfigure}[b]{.45\linewidth}
\includegraphics[width=\linewidth]{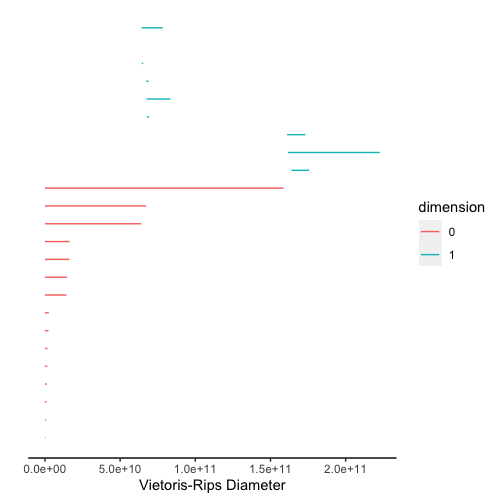}
\caption{Barcode of $X_3$.}\label{figX3}
\end{subfigure}
\caption{Barcodes of the first example.}
\label{fig_ex1}
\end{figure}

In the barcode of $X_0$ (see Fig. \ref{figX0}), we observe that the barcodes of dimension $0$ (the blue barcodes) have exactly $13$ bars; so there are $13$ distinct communities in $X_0$. For the barcodes of $X_1$, $X_2$, $X_3$ (see Fig. \ref{figX1}, \ref{figX2}, \ref{figX3}), we note that all barcodes of dimension $0$ have exactly $15$ bars (which is stabilized), and thus since $X_3$ is the last finite metric space in the filtration
$$X_0 \subset X_1 \subset X_2 \subset X_3,$$
we deduce that there are exactly $15$ distinct communities. This result agrees with the one in Subsection \ref{SubS-patterns}. 

On the other hand, note that in the barcode of dimension $0$ of $X_0$, there are $13$ bars, and in the barcode of dimension $0$ of $X_1$, there are $15$ bars. Since $X_0 = \{1, \ldots, 100\}$, and $X_1 = \{1, \ldots, 300\}$, we conclude that out of $15$ distinct communities in $X_3$, $13$ of them are communities of vertices in $X_0$, and $2$ of them belong to the communities in $X_1$. 

Note that one can also study barcodes of dimension $1$ of the metric spaces $X_k$, and observe that the barcodes of dimension $1$ of $X_1$, $X_2$ and $X_3$ have exactly $5$ important bars, and the remaining bars are \textit{noises}. All these $5$ bars have similar patterns although the number of vertices in $X_k$ changes when $k$ varies from $0$ to $3$. Using persistent homology, one can also realize geometric properties of each $X_k$, for example, how \textit{connected} these spaces are.

For the second example, we choose the $8$ integers $\alpha_2, \ldots, \alpha_9$ to be $3$, $4$, $5$, $7$, $11$, $13$, $17$, $19$, respectively, and the prime number $\beta=23$. We randomly select $700$ integers from $\{1, \ldots, 1000\}$ and sorted in increasing order, denoted as $Y_3$. Let $h=3$, for each $0 \le k < h$, set $Y_k$ to be the first $(2k+1)100$ integers in $Y_3$. Thus we have a filtration of metric spaces
$$Y_0 \subset Y_1 \subset Y_2 \subset Y_3.$$

\begin{figure}
\centering
\begin{subfigure}[b]{.45\linewidth}
\includegraphics[width=\linewidth]{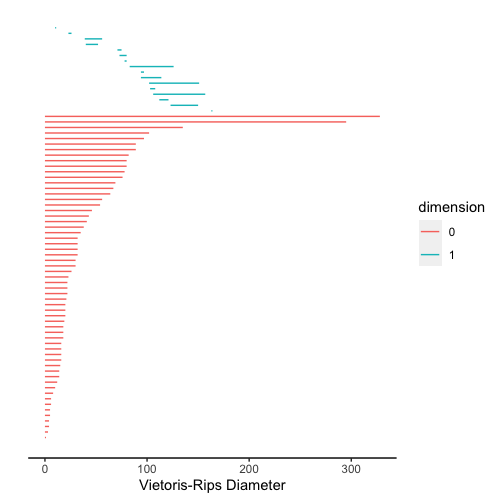}
\caption{Barcode of $Y_0$.}\label{figY0}
\end{subfigure}
\begin{subfigure}[b]{.45\linewidth}
\includegraphics[width=\linewidth]{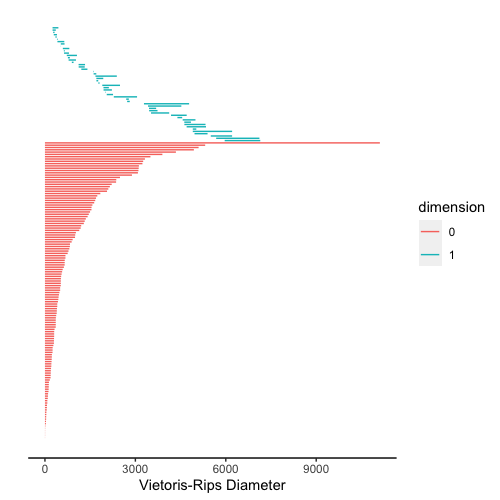}
\caption{Barcode of $Y_1$.}\label{figY1}
\end{subfigure}

\begin{subfigure}[b]{.45\linewidth}
\includegraphics[width=\linewidth]{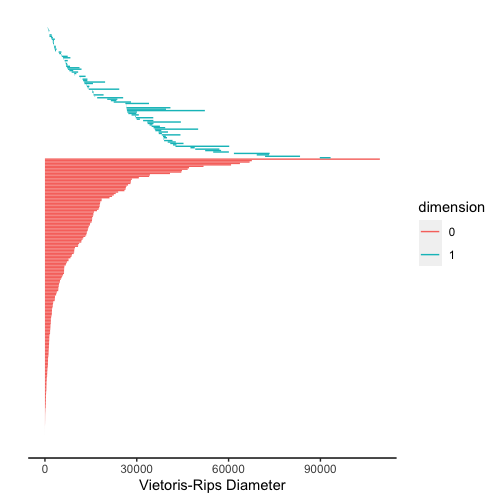}
\caption{Barcode of $Y_2$.}\label{figY2}
\end{subfigure}
\begin{subfigure}[b]{.45\linewidth}
\includegraphics[width=\linewidth]{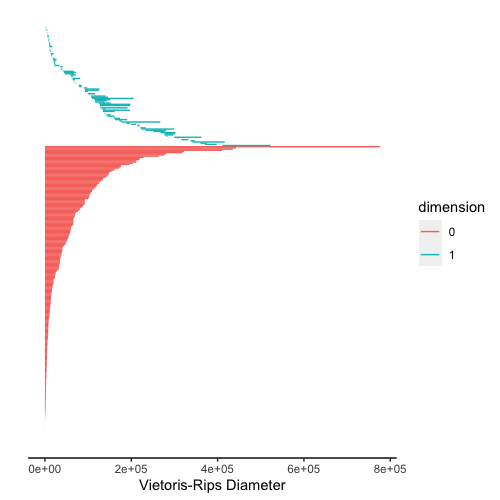}
\caption{Barcode of $Y_3$.}\label{figY3}
\end{subfigure}
\caption{Barcodes of the second example.}
\label{fig_ex2}
\end{figure}

Fig. \ref{fig_ex2} presents the 4 barcodes, each of which corresponds to one $Y_k$. Note that the barcodes of $Y_0$ to $Y_3$ have similar patterns in both of dimension $0$ and dimension $1$. As the number of vertices increases, the barcodes become stabilized.

An important remark is that when comparing the barcodes of the $X$ and $Y$, for example, in dimension $0$, the barcodes of the $Y$ is very connected, which indicates that communities in $Y$ are closely related to each other. This can be seen by observing that each integer can fall into congruence classes of different moduli $3$, $4$, $5$, $7$, $11$, $13$, $17$, $19$. In contrast, in order to define hyperedges of $X$ in the first example, the conditions depend on certain congruence classes modulo $3$, and thus the communities of $X$ are decomposed into distinct connected components which are related to congruence classes modulo $3$ in some way. 

\subsection{Hypergraph-based learning using the metric $\cD$ and nearest neighborhoods}

In hypergraph-based learning, for a given hypergraph, the aim is to find the correct labels for the unlabeled vertices of the test set in the hypergraph under the assumption that one knows the correct labels for the training set. In this subsection, we introduce a modification of the nearest neighbors methods to learn the objective function for a hypergraph. (for the classical nearest neighbors methods, see, for example, in \cite{HTF}.)

Let $\{X, \cP\}_{\Gamma}$ be a hypergraph equipped with properties $\cP$. We equipped $X$ with the metric $\cD$ in Subsection \ref{SubS-Main-Metric} so that $X$ becomes a finite metric space under the metric $\cD$. Suppose that the set of vertices in $X$ consists of $a_1, \ldots, a_n$. Let $f : \cV(X) \to \{-1, 1\}$ be the objective function of labels to be learned such that it sends each vertex to exactly one of the values $-1$ or $1$.  The values of $f$ are also called \textit{signs} of vertices. Let $T =\{(\alpha_i, \beta_i) \; | \; 1 \le i \le m\}$ for some positive integer $1 \le m < n$. Here the $\alpha_i$ are vertices in $X$, and $\beta_i \in \{-1, 1\}$ are the correct label of $\alpha_i$, i.e., $\beta_i = f(\alpha_i)$ for each $1 \le i \le m$. Our aim is to find all values of $\cV(X) \setminus \{\alpha_1, \ldots, \alpha_m\}$ under the objective function $f$, based on the training set $T$. The set $\cV(X) \setminus \{\alpha_1, \ldots, \alpha_m\}$ is called the \textit{test set}. For this purpose, we use the modified nearest neighbors to find a \textit{predictive  model } $f_{\text{NN}}$ for $f$. Fix a positive integer $k \ge 1$. For each vertex $a$ in $X$, we define the following two sets attached to $k$, denoted as $\text{kNN}_1(a)$ and $\text{kNN}_{\text{all}}(a)$ as follows.
\begin{itemize}

\item [(i)] $\text{kNN}_1(a)$ is the set of $k$-th \textit{nearest neighbors} of $a$ in the training set $T$ according to the metric $\cD$. Note that if there are more than one vertex, say $x, y$ in $T$ such that $\cD(a, x) = \cD(a, y)$ and $x, y$ are $k$-th nearest neighbors, then one picks up randomly exactly one such vertex to include in $\text{kNN}_1(a)$.

\item [(ii)] $\text{kNN}_{\text{all}}(a)$ is the set of $k$-th \textit{nearest neighbors} of $a$ in the training set $T$ according to the metric $\cD$. Note that in this set, one includes all vertices $x$ in $T$ such that $x$ is a $k$-th nearest neighbor of $a$. 

\end{itemize}

Using the above two sets $\text{kNN}_1(\cdot)$ and $\text{kNN}_{\text{all}}(\cdot)$, we propose two predictive models for $f$, denoted as $f_{\text{kNN}_1}$ and $f_{\text{kNN}_{\text{all}}}$, respectively. We define
\begin{itemize}

\item [(i)] $f_{\text{kNN}_1}(a) = \text{sign}\left(\sum_{\alpha \in \text{kNN}_1(a)} f(\alpha) \right)$ for each $a$ in the test set.

\item [(ii)] $f_{\text{kNN}_{\text{all}}}(a)= \text{sign}\left(\sum_{\alpha \in \text{kNN}_{\text{all}}(a)} f(\alpha) \right)$ for each $a$ in the test set.

\end{itemize}
Here the sign function is defined by
\begin{align*}
\text{sign}(a) =
\begin{cases}
1 \; \; &\text{if $a \ge 0$} \\
-1 \; \; &\text{if $a < 0$}
\end{cases}
\end{align*}

We illustrate our method by testing on the hypergraph datasets defined in Subsections \ref{example1} and \ref{example2}. Here we define the objective function $f: \cV(X) \to \{-1, 1\}$ as follows: $f(a) = 1$ if $a \equiv 0, 1 \pmod{3}$, and $f(a) = -1$ if $a \equiv -1 \pmod{3}$.

\begin{figure}[t]
\centerline{\includegraphics[scale=0.6]{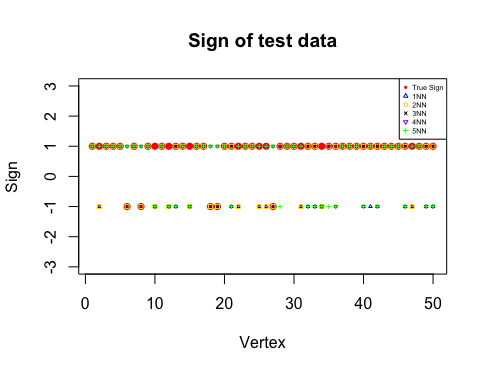}}
\caption{Predicted and true signs of test data for ${\text{kNN}_{\text{all}}}$.}
\label{fig-signs}
\end{figure}

Table \ref{tab1} contains the results of $\text{kNN}$ using the congruence hypergraph defined in the first example, and we set $X_{2000} = \{1, \ldots, 2000\}$. The value of $k$ for $\text{kNN}$ are set to be $1$ to $5$. In each time, we randomly select 70\% vertices from $X$ to be the training set, and we repeat the computation $10$ times for each $k$. Each element in the table presents an error rate which is computed by the percentage of incorrect predictions. According to the average error rates in Table \ref{tab1}, we obtain the smallest average error rate $0.2841$ at $k=3$ when using $\text{kNN}_{\text{all}}$ method and $0.2641$ at $k=2$ when using $\text{kNN}_1$ method. Figure \ref{fig-signs} presents the curve comparison for the predicted and true signs for the method ${\text{kNN}_{\text{all}}}$. In this figure, the error rates of $f_{\text{kNN}_{\text{all}}}$ are from the ninth row in Table \ref{tab1}. According to the figure, most of vertices with label 1 are predicted correctly. One of the reasons that cause this result is that the number of vertices with positive sign are much larger then the number of vertices with negative sign according to the way we define the objective function.

Table \ref{tab2} contains the results of $\text{kNN}$ using the congruence hypergraph defined in the second example. We randomly select $5000$ vertices from $\{1, \ldots, 8000\}$, the values of $\{\alpha_2, \ldots, \alpha_9\}$ and $\beta$ are the same as described in section \ref{homology using metric}. Using the $\text{kNN}_{\text{all}}$ method, the smallest average error rate is $0.3463$ at $k=5$. Using the $\text{kNN}_1$ method, the smallest average error rate is $0.3419$ at $k=2$.  According to the results in Table \ref{tab1} and \ref{tab2}, the $\text{kNN}_{\text{all}}$ method performs slightly better then $\text{kNN}_1$.

\begin{table}[t]
\caption{Error rates of KNN using the first example}
\begin{center}
\begin{tabular}{|c|c|c|c|c|c|c|c|c|c|c|}
\hline
&\multicolumn{5}{|c|}{\textbf{Error rate of $f_{\text{kNN}_{\text{all}}}$}}&\multicolumn{5}{|c|}{\textbf{Error rate of $f_{\text{kNN}_1}$}} \\
\hline
\textbf{ Error rate} & \textbf{K=1}& \textbf{K=2}& \textbf{K=3} & \textbf{K=4} & \textbf{K=5} &\textbf{K=1}& \textbf{K=2}& \textbf{K=3} & \textbf{K=4} & \textbf{K=5}\  \\
\hline
1& 0.3263   & 0.3200   & 0.3200  &  0.3200  &  0.3200 & 0.4762 &   0.2700 &   0.4012 &   0.2800 &   0.3775\\
\hline
2& 0.3187&    0.4613 &   0.1887 &    0.3925  &  0.4463 & 0.4712   & 0.2837  &  0.4712 &   0.2700 &   0.4225\\
\hline
3&0.2800  &  0.3163 &   0.2987 &   0.3888 &   0.3050 & 0.5100  &  0.2913 &   0.4975 &   0.3337 &   0.4525\\
\hline
4& 0.3762   & 0.3225  &  0.3313 &   0.1850  &  0.1775 & 0.4087 &   0.2213  &  0.2813  &  0.1900 &    0.3075\\
\hline
5& 0.3013  &  0.2925&    0.2925 &   0.2925  &  0.3850 & 0.4225 &   0.2650  &  0.3812 &   0.2450  &  0.3800\\
\hline
6& 0.4400 &   0.2538  &  0.3938  &  0.3775  &  0.4150 & 0.4437   & 0.2163  &  0.3888 &   0.2875 &   0.3938\\
\hline
7& 0.1562 &   0.2675&    0.2675 &   0.4225  &  0.3063 & 0.4587  &  0.2562 &  0.3975 &   0.2850 &   0.3800\\
\hline
8&  0.3137 &   0.2762  &  0.2712 &   0.2712 &   0.2850 & 0.4625  &  0.2450  &  0.4050  &  0.2750  &  0.3938\\
\hline
9& 0.4250 &   0.2638 &   0.1125 &   0.2438 &   0.2825 &0.4675  &  0.2312  &  0.4663  &  0.2937    &0.4287\\
\hline
10&  0.4350 &   0.3900 &   0.3650  &  0.3275 &   0.3550&  0.5075&    0.2688  &  0.4525  &  0.3363  &  0.4300\\
\hline
Average& 0.3372& 0.3164& 0.2841& 0.3221& 0.3278 & 0.4626& 0.2641& 0.4098& 0.2866& 0.3946 \\
\hline
\end{tabular}
\label{tab1}
\end{center}
\end{table}

\begin{table}[t]
\caption{Error rates of KNN using the second example}
\begin{center}
\begin{tabular}{|c|c|c|c|c|c|c|c|c|c|c|}
\hline
&\multicolumn{5}{|c|}{\textbf{Error rate of $f_{\text{kNN}_{\text{all}}}$}}&\multicolumn{5}{|c|}{\textbf{Error rate of $f_{\text{kNN}_1}$}} \\
\hline
\textbf{ Error rate} & \textbf{K=1}& \textbf{K=2}& \textbf{K=3} & \textbf{K=4} & \textbf{K=5} &\textbf{K=1}& \textbf{K=2}& \textbf{K=3} & \textbf{K=4} & \textbf{K=5}\  \\
\hline
1& 0.3527 & 0.3447  & 0.3447&0.3440 &0.3433 & 0.3713 & 0.3347&0.3687 & 0.3260 & 0.3500\\
\hline
2& 0.3567 & 0.3413 & 0.3487 & 0.3347 & 0.3360 & 0.3493 & 0.3367 & 0.3660 & 0.3453 & 0.3527 \\
\hline
3& 0.3793 & 0.3693 & 0.3633 & 0.3433 & 0.3480 & 0.3913 & 0.3473 & 0.3933 & 0.3413 & 0.3540\\
\hline
4& 0.3733 & 0.3580 & 0.3540 &0.3500 & 0.3460& 0.3707 & 0.3487 &0.3813 &0.3433 &0.3580 \\
\hline
5& 0.4000 & 0.3660 & 0.3607 & 0.3627 & 0.3593 & 0.3960 & 0.3540 & 0.3807 &0.3573 & 0.3780\\
\hline
6& 0.3853 & 0.3580 & 0.3653 & 0.3453 & 0.3460 & 0.3873 & 0.3380 & 0.3893 & 0.3447 & 0.3680\\
\hline
7& 0.3520 & 0.3347 & 0.3373 & 0.3493 & 0.3453 & 0.3680 & 0.3333 & 0.3720 & 0.3420 & 0.3520\\
\hline
8&  0.3773 & 0.3587 & 0.3687 & 0.3547 & 0.3540 & 0.3627 & 0.3513 & 0.3847 & 0.3500 & 0.3567\\
\hline
9& 0.3707 & 0.3500 & 0.3653 & 0.3453 & 0.3467 & 0.3807 & 0.3527 & 0.3800 & 0.3493 & 0.3640\\
\hline
10&  0.3633 & 0.3427 & 0.3473 & 0.3360 & 0.3380 & 0.3707 & 0.3227 & 0.3613 & 0.3333 & 0.3540\\
\hline
Average& 0.3715 & 0.3523 & 0.3555 &0.3465 & 0.3463& 0.3748 & 0.3419 & 0.3777& 0.3460 &0.3587 \\
\hline
\end{tabular}
\label{tab2}
\end{center}
\end{table}

\section{Conclusions}

Our main contributions in this paper can be summarized as follows:
\begin{itemize}

\item[(i)] Introducing a natural metric space structure modulo certain equivalence relations on a general hypergraph data which bears a resemblance to a usual metric space structure;

\item [(ii)] Using the metric space modulo certain equivalence relation structure introduced, we emphasize that this topological space structure on hypegraphs is very natural and suitable for studying several problems in machine learning;

\item [(iii)] Proposing a distance matrix approach using the metric space structure introduced in this paper to study community detection problem in hypegraphs;

\item [(iv)] Proposing a modified homology-based learning to study topological structures of hypergraphs which in turn can be used to detect persistent homological features; this method can also be used to study community detection problem;

\item [(v)] Proposing modified nearest neighbors methods for studying sign prediction problem on general hypergraph data; such methods have advantages that they can be applied even to hypergraphs which do not contain an embedding into a Euclidean space, or do not carry a Euclidean space structure.

\item [(vi)] One of our main contributions is to propose a new way to construct hypergraph data which are very easy to implement and test theories from machine learning which we coin the term ``\textit{congruence hypergraph data}''.

\item [(vii)] Experimental analysis are performed on congruence hypergraph data which are simulated by our methods.
\end{itemize}

\section{Acknowledgements}

Lizhen Lin would like to acknowledge the support of NSF grant DMS CAREER 1654579.

\section{Appendix}

In this Appendix, we give a proof of Theorem \ref{main-thm1}

For the sake of simplicity, let $\alpha_i = \cN(a_i)$ for each $1 \le i \le n$. 

Suppose that $\cD(\alpha_i, \alpha_j) = 0$ for some $1 \le i, j \le n$. By definition, we know that
\begin{align*}
\cD(\alpha_i, \alpha_j) = \sum_{P \in \cP}\sum_{m = 1}^{\ell}|\fC^m_P(a_i) - \fC^m_P(a_j)| = 0,
\end{align*}
which implies that 
\begin{align*}
\fC^m_P(a_i) - \fC^m_P(a_j) = 0
\end{align*}
for all $m \ge 1$ and $P \in \cP$. Thus $\fC^m_P(a_i) = \fC^m_P(a_j)$ for all $m \ge 1$ and $P \in \cP$, and hence $\alpha_i = \cN(a_i) \cong \alpha_j = \cN(a_j)$. 

It is obvious that $\cD(\alpha_i, \alpha_j) = \cD(\alpha_j, \alpha_i)$ for all $1 \le i, j \le n$, which proves that $\cD$ is symmetric.

We now show that $\cD$ satisfies the triangle inequality. Indeed, we see that
\begin{align*}
 &\cD(\alpha_i, \alpha_j) = \sum_{P \in \cP}\sum_{m = 1}^{\ell}|\fC^m_P(a_i) - \fC^m_P(a_j)| \\
 &= \sum_{P \in \cP}\sum_{m = 1}^{\ell}|(\fC^m_P(a_i) - \fC^m_P(a_k)) + (\fC^m_P(a_k) - \fC^m_Q(a_j)| \\
 &\le \sum_{P \in \cP}\sum_{m = 1}^{\ell}|\fC^m_P(a_i) - \fC^m_P(a_k)| + \sum_{P \in \cP}\sum_{m = 1}^{\ell}|\fC^m_P(a_k) - \fC^m_P(a_j)| \\
 &=  \cD(\alpha_i, \alpha_k) +  \cD(\alpha_k, \alpha_j)
 \end{align*}
 for any $1 \le i, j, k \le n$. Thus $\cD$ satisfies the triangle inequality, and therefore $\cD$ is a metric on $X$ modulo the equivalent relation $\cong$.

\end{document}